\def\stwo{{s^\text{(2)}}}
\def\stwot{{s^{\text{(2)}}_\text{t}}}

\def\grt{{g}_\text{tf}(r)}

\def\exsp{\overline{s^{\text{ex}}_\text{t}}}

\def\Vtp{{\phi _\mathrm{tf}}(r)}

\documentclass[superscriptaddress,prb,preprint,amsmath,amssymb]{revtex4}

\usepackage{graphicx,natbib}

\usepackage{calc,graphicx,float,amsmath}

\usepackage[paper=letterpaper,
            includefoot,
            marginparsep=.05in,       
            margin=0.75in,               
            includemp]{geometry} 

\usepackage{bm}  
\begin{document}

\title{Enhancing tracer diffusivity by tuning interparticle interactions and coordination shell structure}

\author{James Carmer} 
\affiliation{Department of Chemical Engineering,
  The University of Texas at Austin, Austin, TX 78712.}

\author{Gaurav Goel} 
\affiliation{Department of Chemical Engineering, 
	Indian Institute of Technology, Delhi, Hauz Khas, New Delhi 110016.}
 
\author{Mark J. Pond} 
\affiliation{Department of Chemical Engineering,
  The University of Texas at Austin, Austin, TX 78712.}

\author{Jeffrey R. Errington} 
\affiliation{Department of Chemical and Biological Engineering,
  University at Buffalo, The State University of New York, Buffalo,
  New York 14260-4200, USA}

\author{Thomas M. Truskett} \email{truskett@che.utexas.edu}
\thanks{Corresponding Author} {}
\affiliation{Department of Chemical Engineering, The University of Texas
  at Austin, Austin, TX 78712.}
  
\footnotetext{\text{$^{\ddag}$~These authors contributed equally to this work.}}

\begin{abstract}
This study uses a combination of stochastic optimization, statistical mechanical theory,  and molecular simulation to test the extent to which the long-time dynamics of a \textit{single} tracer particle can be enhanced by rationally modifying its interactions--and hence static correlations--with the other particles of a dense fluid. Specifically, a simulated annealing strategy is introduced that, when coupled with test-particle calculations from an accurate density functional theory, finds interactions that maximize either the tracer's partial molar excess entropy or a related pair-correlation measure (i.e., two quantities known to correlate with tracer diffusivity in other contexts). The  optimized interactions have soft, Yukawa-like repulsions, which extend beyond the hard-sphere interaction and disrupt the coordination-shell cage structure surrounding the tracer. Molecular and Brownian dynamics simulations find that tracers with these additional soft repulsions can diffuse more than three times faster than bare hard spheres  in a moderately supercooled fluid, despite the fact that the former appear considerably larger than the latter by conventional definitions of particle size.
\end{abstract}
\maketitle

\section {Introduction\label{sec:introduction}}

Diffusion of a tracer particle through a complex fluid is an essential process for many chemical, materials, and biological systems.  The
rate at which it occurs--characterized, e.g., by the 
corresponding long-time diffusion coefficient--has practical 
implications for 
technological developments in drug delivery, catalysis, 
materials synthesis, and
separations.\cite{Cussler2009Diffusion:mass,Dhont1996introductiontodynamics,Saltzman2001Drugdelivery:,Pluen2001Roleoftumor-host}
Experimental studies clearly show that there is an
intimate
connection between tracer diffusivity and the interparticle
interactions present in a system.\cite{0953-8984-5-34B-021,Imhof1994comparisonbetweenlong-time,Kluijtmans1998Self-diffusionofcharged}
Furthermore, approximate microscopic approaches for predicting dynamics, 
such as
kinetic,\cite{Lowen1993Long-timeself-diffusioncoefficient,Feng2004Diagrammatickinetictheory}
generalized  Langevin,\cite{Medina-Noyola1987TheGeneralizedLangevin,Hernandez-Contreras1996Generaltheoryof,Juarez-Maldonado2008Theoryofdynamic}
  and mode-coupling\cite{Nagele1998Tracer-diffusionincolloidal,Krakoviack2005Liquid-GlassTransitionof,Viehman2008Theoryofgelation,Viehman2008DynamicsofTracer}
theories, have recently provided some important insights into this link. Nonetheless, a comprehensive theoretical framework--or even a
clear conceptual picture--for understanding and predicting {\em how} tracer diffusivity should depend on the 
interparticle interactions is still lacking. As a result, basic questions about how to engineer complex fluids with desired dynamic properties remain challenging to address, especially for cases where nontrivial constraints on particle properties (size, chemistry, etc.) and thermodynamic conditions (e.g., particle concentration, composition, etc.) must  be satisfied. 

In this paper, we present theoretical results that shed new light on basic physics relevant to 
designing single-particle dynamics of complex fluids.  In particular, we explore--within a model system--the  extent to which one can rationally modify the tracer diffusivity of a particle by tuning how it interacts with the other particles in the system.  
For colloids or nanoparticles suspended in solvent, this is particularly interesting because it is
often possible to systematically modify their
effective interparticle interactions by varying solvent quality
(e.g., by changing composition of a third component such as a
surfactant or salt), by introducing a 
depleting agent (e.g., a nonadsorbing polymer), by chemically
attaching or physically adsorbing molecules onto the
particles, or by tuning of electric or magnetic interactions via an
external
field.\cite{Blees1996Self-DiffusionofCharged,Blaaderen1992Long-timeself-diffusionof,Koenderink2002Rotationaldynamicsof,0022-3727-36-13-201,Pantina2008Micromechanicsandcontact,Erb2009Magneticassemblyof}

As a starting point for our analysis, we consider a single hard-sphere tracer particle in a dense fluid of other hard spheres. The question we address here is simply, which types of 
isotropic  pairwise potentials between the tagged tracer particle and the neighboring particles--when added to the bare hard-sphere interaction--significantly increase the tracer's long-time diffusion coefficient?  

At first
glance, the idea that adding a  contribution
to the tracer particle's hard-sphere interaction could significantly \textit{increase} its long-time mobility may not be intuitively obvious.
The naive expectation is that adding an attractive interaction would 
slow down the dynamics of a single tracer 
because it would energetically stabilize the surrounding
``cage'' structure formed by its nearest neighbors.  On the other hand, adding
a repulsive interaction to the pair potential 
might likewise be expected to decrease the mobility
of the tracer because it would effectively increase its size. Based on these  arguments, one  might hypothesize that the bare hard-sphere
interaction is optimal for tracer diffusivity.  
Interestingly, as we discuss in detail below, adding certain types of softer, repulsive interactions to the
tracer's hard core can result in a pronounced increase in its mobility. The key to achieving the enhanced diffusivity appears to be  finding interactions that disrupt the coordination-shell structure surrounding the tracer without significantly increasing the tracer's effective size. 

How can one discover the mathematical forms of such mobility enhancing interactions in a reasonably efficient and systematic way?
The strategy we pursue here is to first identify 
structural or thermodynamic
quantities, relatable to interparticle interactions through
equilibrium statistical mechanics, that correlate strongly with tracer 
diffusivity in other related systems.  Once an appropriate static property has been identified, we employ
a variational approach that uses either computer simulations or
liquid-state theory to find the types of interparticle interactions
that increase (perhaps even optimize) the static property, subject to relevant realizability constraints.  Finally, 
we perform molecular or Brownian dynamics simulations
to determine whether the interactions do in fact result in tracer-particle long-time diffusivities significantly higher than those of the  reference 
hard-sphere particles.   

The specific strategy we adopt here is motivated by the observation from molecular simulations and experiments that the excess entropy
(relative to an ideal gas in the same volume) $s^{\text{ex}}$, as well as its associated two-particle
approximation $\stwo$, can be used to
semi-quantitatively predict a number of nontrivial effects that 
not only temperature and particle
concentration\cite{Rosenfeld1977Relationbetweentransport,Rosenfeld1999quasi-universalscalinglaw,Dzugutov1996universalscalinglaw,Mittal2007RelationshipsbetweenSelf-Diffusivity,Sharma2006Entropydiffusivityand,Agarwal2007Ionicmeltswith,Yan2008Relationofwater,PondGCBinary,GeneralizedRosenfeld2009,PondBrownian2011, Abramson2007Viscosityofwater,Abramson2008Viscosityofnitrogen,Abramson2009ViscosityCO2,hoover1991computational,DyreIsomorphs2008}--but also confinement and other external fields\cite{Mittal2007Doesconfininghard-sphere,Goel2008Tuningdensityprofiles,Mittal2008LayeringandPosition,Mittal2006ThermodynamicsPredictsHow,Mittal2007RelationshipsbetweenSelf-Diffusivity,Mittal2007Confinemententropyand,GoelAvailableSpace2009,PhysRevE.82.041201}--have on the transport coefficients of  equilibrium and moderately supercooled fluids. Excess
entropy is a static measure that characterizes the reduction in the number
of states available to a system due to interparticle
correlations. Although a rigorous link between excess entropy and dynamics that can explain all of the aforementioned observations is still lacking, the connection is now well understood for low-density gases,\cite{Rosenfeld1999quasi-universalscalinglaw} fluids with inverse-power-law pair potentials,\cite{hoover1991computational} and fluids that exhibit so-called ``isomorphs''.\cite{DyreIsomorphs2008} It is also physically reasonable to expect that the collisional processes important for dynamics correlate with structural measures that track the strength of the static interparticle correlations (at least for conditions not too close to the glass transition).

There are two excess-entropy based quantities that are natural to investigate when the focus is on the structure and dynamics of a single tracer particle:  (i) the two-body
contribution to excess entropy, $\stwot$, arising from the partial radial
distribution function between the tracer and its neighbors, and, (ii)
the partial excess molar entropy
of the tracer $\exsp$. Both have been found in molecular simulations to correlate with the long-time dynamics of particles in mixtures.\cite{PondGCBinary,Mittal2007Confinemententropyand}

We organize the balance of the paper as follows.  In Sec.~\ref{theory_methods}, we 
provide information about the model system that we study, and we 
discuss how to compute $\stwot$ and
$\exsp$ for this system from statistical mechanics.   We then discuss simple
variational schemes for obtaining tracer-fluid
potentials which result in significantly higher values of $\stwot$ and
$\exsp$ as compared to those associated with 
the bare, hard-sphere tracer.  We also present details concerning 
how we compute the corresponding tracer diffusivities 
via molecular (and, in a few cases for comparison, Brownian) dynamics simulations.  In Sec.~\ref{sec:RandD},
we show the link between tracer diffusivity and the two 
aforementioned static measures for hard-sphere tracer particles of
various size.  We then demonstrate that the variational calculations discussed
above produce soft, Yukawa-like repulsive potentials that extend beyond  the
bare hard-sphere interaction and disrupt the coordination-shell cage structure surrounding the tracer.  Particles with the
additional soft potentials can have tracer diffusivities more than three
times higher than those of the bare hard-sphere particles in the moderately supercooled fluid.  This is
striking, especially given that the former would be considered ``larger'' than the latter by conventional measures (e.g., Barker-Henderson radius, second-virial
coefficient, partial molar volume, and solubility).  Finally, we present calculations using a model tracer particle with hard-core plus Yukawa interactions that  illustrate how the optimized interactions presented in this study strike a  balance between being strong enough to reduce local coordination-shell structure that impedes diffusivity, but still soft enough to avoid typical decreases in mobility due to larger particle size.  

\section {Theory and Methods \label{theory_methods}}

\subsection{Model Interactions}
\label{sec:md_simulations}
Consider a single tracer (t) particle  in a bath
of hard-sphere-like fluid (f) particles.  We model the hard cores of all particles
in the system via pair potentials of the following 
Weeks-Chandler-Andersen (WCA) form,\cite{Chandler1983VanderWaals}
\begin{equation}
  \label{eq:WCA}
  \phi^\text{WCA}(r;d,\epsilon)= \begin{cases}
  4\epsilon([d/r]^{48}-[d/r]^{24})+\epsilon& r < 2^{1/24}d\\
    0& r \ge 2^{1/24}d\\ \end{cases}
\end{equation}The quantities $d$ and $\epsilon$ characterize the usual length and energy
scales of the WCA interaction, respectively.  
The fluid particles surrounding the tracer have a nominal diameter $\sigma_\text{}$  and interact solely via this 
potential,
\begin{equation}
  \label{eq:WCAff}
  \phi_\mathrm{ff} (r)=\phi^\text{WCA}(r;\sigma,\epsilon)
\end{equation}The pair interaction between the tracer particle and the surrounding fluid
particles is given by:
\begin{equation}
  \label{eq:WCA_tf}
   \Vtp= \begin{cases}
\phi^\text{WCA}(r;\sigma_\text{tf},\epsilon)& r<r_\text{0}\\

  \phi^\text{WCA}(r;\sigma_\text{tf},\epsilon)+\phi_\text{0}(r)& r \ge
  r_\text{0}\\ \end{cases}
\end{equation}
Here, $\sigma_\text{tf}=(\sigma_\text{t}+\sigma)/2$, and
$\sigma_\text{t}$ is the nominal diameter of the tracer particle.  
The additional interaction, $\phi_\text{0}(r)$, is the
contribution that  can be ``tuned'' in order to maximize $\stwot$ or  $\exsp$, and --as we show-- increase the dynamics of
the tracer particle.  The procedure for
computing $\phi_\text{0}(r)$ is discussed in
Sec.~\ref{sec:variational_scheme}~and~\ref{sec:interactions_faster_diffusion_experimental}. To
ensure that the tuning procedure  does not impact the size of the exclusion region that constitutes the bare hard-core interaction, we define $r_\text{0}$ such that $g_\text{tf}(r) \ge 5\times10^{-3}$ for $r \ge r_0$ (and $g_\text{tf}(r) < 5\times10^{-3}$ for $r < r_0$), where $\grt$ is the partial radial distribution
function (PRDF) between the tracer and the other fluid particles. For the range of parameters studied in this work, this leads to  
$r_\text{0} = (0.97\text{--}0.98) \sigma_\text{tf}$, and
$\phi^\text{WCA}(r_\text{0};\sigma_\text{tf},\epsilon)=5$--$8
k_\text{B}T$. 

\subsection{Calculation of $\stwot$ and $\exsp$}
\label{sec:calc_entropy}
Consider a binary fluid mixture of $N_{\rm t}$ tagged tracer particles and $N$ other particles in the infinite dilution limit of the former [i.e., vanishing tracer-particle mole fraction $x_{\rm t}=N_{\rm t}/(N_{\rm t}+N) \rightarrow 0$]. The two-body contribution to the excess entropy of the fluid 
arising from the tracer-fluid
positional correlations in this limit is given by\cite{Baranyai1989DirectEntropyCalculation,mountain1971EntropyAndMolecular}
\begin{equation}
  \label{eq:s2tracer}
  \stwot=-\frac{\rho k_\text{B}}{2}\int \{\grt \ln \grt -\grt+1\}d\mathbf{r}
\end{equation}
where $\rho=N/V$, $V$ is the volume,  and $k_\text{B}$ is
Boltzmann's constant.  

More generally, the  partial molar excess entropy associated with the tracer particle is given by 
\begin{eqnarray}
\exsp=\left[\left(\frac{\partial S}{\partial
    N_\mathrm{t}}\right)_{T,P,N_\mathrm{}}-\left(\frac{\partial S^\text{ig}}{\partial
    N_\mathrm{t}}\right)_{T,P,N}\right] \label{eq:def_pexs}
\end{eqnarray}where $S$, $T$, and $P$ are the entropy, temperature, and pressure of the fluid, respectively. $S^{\rm ig}$ is the corresponding entropy of an interaction-free (i.e., ideal gas) version of the system with the same values of  $V$, $N_{\rm t}$, and $N$. Using standard thermodynamic relations, one can recast $\exsp$ as follows, \begin{equation}
\label{eq:solute_muex}
  \exsp/k_\text{B}= \beta \overline
  {e{^{\text{ex}}_\text{t}}}+ (Z-1)\rho\overline {v}_\text{t}-\beta \mu ^{\text{ex}}_\text{t} 
\end{equation}
Here $\beta=(k_{\rm B}T)^{-1}$, and $Z=\beta P/\rho$ is the compressibility factor of the fluid. The quantities $\overline
  {{e}{^{\text{ex}}_\text{t}}}$, $\overline {v}_\text{t}$, and $\mu ^{\text{ex}}_\text{t}$ represent the tracer particle's excess energy, partial molar volume, and excess chemical potential, respectively. 

For the infinitely dilute tracer particle in a fluid of hard-spheres considered here, each term on the right-hand side of 
eq.~\ref{eq:solute_muex} can be simplified considerably. 
Specifically, the partial molar excess energy of the tracer is given by
\begin{equation}
\label{eq:tracer_energy}
  \overline {e{^{\text{ex}}_\text{t}}}=  \rho \int \grt \Vtp d\mathbf{r}
\end{equation}

The partial molar volume for the tracer can be calculated using the 
Kirkwood-Buff relationship:\cite{Kirkwood1951StatisticalMechanicalTheory}
\begin{equation}
\label{eq:tracer_vp}
\overline {v}{_\text{t}} =  \kappa k_\text{B}T + \int [1-\grt] d\mathbf{r}
\end{equation}
where $\kappa$ is the isothermal compressibility of the pure hard-sphere
fluid, which we estimate here using
the Carnahan-Starling equation of state.\cite{Carnahan1969EquationOfState}
Note the integral in eq.~\ref{eq:tracer_vp} converges slowly and hence its numerical value is
sensitive to the truncation point. However, as discussed elsewhere,\cite{Lazaridis1998InhomogeneousFluidApproach} 
its value can be accurately estimated by extrapolating the values obtained when integrating out to various maxima and minima of $g_\text{tf}(r)$ (in the range $r=6-10 \sigma$).

Finally, the excess chemical potential of the tracer is equal the reversible 
work of inserting a test tracer particle into the hard-sphere fluid at constant
temperature $T$ and chemical potential $\mu$,\cite{Hendersen1983StatisticalMechanicsOf}
\begin{equation}
\label{eq:tracer_mu}
  \mu{^{\text{ex}}_\text{t}}=  \Omega_\text{t}[\grt;T, \mu] -\Omega(\mu, T)
\end{equation}
Here, $\Omega _\text{t}$ is the grand-potential of the fluid with
the tracer particle fixed at the origin, and $\Omega$ is the
grand-potential of pure fluid without the tracer particle. 
Density functional theories for inhomogeneous
fluids, like the one discussed in Sec.~\ref{sec:DFT}, 
provide an approximate functional relationship for 
$\Omega_\text{t}[\grt;T, \mu]$, while $\Omega(\mu, T)$ 
is a property of
the pure hard-sphere fluid.  In this work, the latter is again estimated from the Carnahan-Starling equation of state.\cite{Carnahan1969EquationOfState}  

A key observation of this section is that, for an infinitely dilute tracer in a hard-sphere fluid, $\stwot$ is a  functional of $\grt$, and $\exsp$  is a functional of $\grt$ and $\Vtp$. Moreover, a basic result of statistical mechanics\cite{Weeks2003ExternalFieldsDensity} is that $\grt$ uniquely determines $\Vtp$--and vice versa--for this problem. As a consequence, one can search out  optimal functions for $\grt$ [or equivalently $\Vtp$] that maximize either $\stwot$ or $\exsp$ subject to suitable realizability constraints.   

\subsection{Density functional theory}
\label{sec:DFT}

Classical density functional theory (DFT)\ of inhomogeneous fluids provides a convenient means for interconverting between $\grt$ and
$\Vtp$. In this work, we use a recent modification\cite{Yu2002Structuresofhard-sphere} of
Rosenfeld's fundamental measure theory
\cite{Rosenfeld1989Free-energymodelinhomogeneous,
Rosenfeld1997Fundamental-measurefree-energydensity}
 which--by construction--reproduces the Carnahan-Starling equation of state for a hard-sphere fluid in the homogeneous limit. For our DFT calculations, we map the fluid of 
hard-sphere-like WCA particles onto ``equivalent'' 
hard spheres of diameter
$\sigma_\text{f}$, an approximation that has
been found to be adequate in other 
contexts.~\cite{Goel2008Tuningdensityprofiles}
The tracer particle, on the other hand, is treated as a fixed 
external potential of the form $\Vtp$ [see eq.~\ref{eq:WCA_tf}].
This ``test-particle'' approach allows us
to use the inhomogeneous one-component version of fundamental measure
theory to compute infinite-dilution partial molar properties of the tracer particle in a homogeneous hard-sphere fluid.~\cite{Roth2000Depletionpotentialin}

The following expression provides a convenient representation for computing $\grt$ within this
framework,
\begin{equation}
  \label{eq:DFT}
   \grt = \frac{1}{\rho \Lambda_\text{b}^3} \exp \left\{\beta[\mu-\Vtp]+c^{(1)}[r,\grt;\mu]\right\}
\end{equation}
where $\Lambda_\text{b}$ is the thermal deBroglie 
wavelength of the fluid particles and $c^{(1)}$
is their one-body direct correlation function in the external field of
the test particle.  The latter can be
expressed as a functional derivative of excess intrinsic Helmholtz
free energy $\cal{F}^{\mathrm{ex}}$ with respect to $\grt$,
\begin{equation}
  \label{eq:c1}
c^{(1)}=-\frac{1}{\rho}\frac{\delta{\cal{F}}^{\mathrm{ex}}[\grt;\mu]}{\delta
  \grt}
\end{equation}
Explicit functional expressions for both $\cal{F}^{\mathrm{ex}}$ and
$\Omega_\text{t}$ of eq.~\ref{eq:tracer_mu} from the
modified fundamental measure theory are presented elsewhere.~\cite{Yu2002Structuresofhard-sphere}

To obtain $\grt$ for a given $\Vtp$, $T$, and $\mu$,
we solve eq.~\ref{eq:DFT} using Picard iterations.\footnote{Weighted densities and convolution integrals were
calculated using the Romberg integration scheme on a mesh of
$0.01\sigma_\text{f}$. To avoid divergence, we mix the input
$\grt_\mathrm{in}^{n}$ and the output $\grt_\mathrm{out}^{n}$
profiles for iteration $n$ to obtain the input
profile for iteration $n+1$, i.e., $\grt_\mathrm{in}^{n+1}  =
\grt_\mathrm{out}^{n} \lambda + \grt_\mathrm{in}^{n}
(1-\lambda)$. We set the mixing parameter $\lambda=0.01$. 
 We continue iterations until the root mean square
 change in $\grt$ is less than $1 \times 10^{-8}$.} 
 Alternatively, for a given $\grt$, $T$, and
$\mu$ (and hence $\rho$), we obtain the corresponding $\Vtp$ by numerically inverting
eq.~\ref{eq:DFT}.

\subsection{Optimizing tracer-fluid interactions}
\label{sec:variational_scheme}

\subsubsection{Optimizing $\stwot$}
\label{sec:opt_stwo}
 
~

From eq.~\ref{eq:s2tracer}, it can easily be seen that a ``flattened" PRDF--i.e., $\grt=\Theta (r-\sigma_\text{tf})$, where
$\Theta (r-\sigma_\text{tf})$ is the Heaviside step function--maximizes
$\stwot$ subject to the hard-core constraint. We insert this  target $\grt$ into the inverse DFT  protocol described in section~\ref{sec:DFT} to determine the corresponding
tracer-fluid interactions~$\Vtp$ for various values of $\sigma_\text{t}/\sigma_\text{}$ and reduced fluid density $\rho \sigma^3$. We also compute the partial molar excess entropy $\exsp_{\rm }$ of this particle with the methods described in
section~~\ref{sec:calc_entropy}.           

\subsubsection{Optimizing $\exsp$}
\label{sec:opt_exsp}

~

To compute tracer-fluid interactions~$\Vtp$ [and corresponding $\grt$]  that minimize $-\exsp/k_{\rm B}$, subject to the hard-core constraint, we use a simulated annealing global optimization algorithm
described in Corana et
al.\cite{Corana1987MinimizingMultimodalFunctions}.  For the optimization algorithm, we 
assume that $\grt$ for $r_0 <r \le r_{\rm a}$ can be approximately described  by the following expression:
\begin{equation}
  \label{eq:grt_fourier}
  \grt=\exp \left [ a_0 + \sum_{m=1}^{35}  a_m\cos \left(\lambda_m\{r-r_0\}\right) +
    \sum_{n=1}^{5} \frac{b_n}{r^{n}}
  \right ]
\end{equation}
 Here, 
$\lambda_m=m \pi /(r_\text{a}-r_\text{0})$, and $r_\text{a}$ is chosen such that $|g_{\rm{tf}}(r_{\rm a})-1|<0.02$ for a single hard-sphere-like tracer particle of diameter $\sigma_{\rm t}$ surrounded by a fluid of hard-sphere-like particles of diameter $\sigma$ with  density $\rho\sigma^3$   .   For $r>r_\text{a}$, 
we adopt a standard asymptotic form for 
$\grt$\cite{Roth2000Depletionpotentialin} and determine its
required coefficients from a least-squares fit to the same hard-sphere tracer's PRDF data.  

The optimization algorithm proceeds as follows. An initial 
coefficient vector [$a_{1},...,a_{35}, b_{1},...,b_{5}$] is chosen by fitting eq.~\ref{eq:grt_fourier} to simulation data for the equilibrium structure surrounding  the aforementioned hard-sphere-like tracer particle. The objective function ($-\exsp_{\rm current}/k_{\rm B}$) for this system is then computed using the method described in section \ref{sec:calc_entropy}. An
initial effective temperature for the simulated annealing algorithm, $k_{\rm B}T_\text{sa}$ ($=1\times10^{-4}$ in this work) is also provided
as input. 

A cycle in the simulated annealing scheme is then carried out 
as follows. (i) A ``trial'' coefficient vector is created by randomly modifying the coefficients of the ``current'' coefficient vector. The magnitudes of the modifications are chosen such that approximately half of trial coefficient vectors are accepted according to the criteria described in the next step.  (ii) If the trial $\grt$ does not satisfy the constraints described in section  \ref{sec:md_simulations}, then it is rejected and step (i) is repeated. Otherwise, it is input into the inverse DFT routine of section \ref{sec:DFT} to
determine $\Vtp$, and the corresponding partial molar excess entropy $\exsp_{\rm trial}$ is computed as described in
section~\ref{sec:DFT}~and~\ref{sec:calc_entropy}, respectively.  If the trial $\grt$ results in   $-\exsp_{\rm trial} < -\exsp_{\rm current}$, then it is accepted and it replaces the current PRDF.  If not, the trial PRDF is accepted with probability $\exp
[(\exsp|_\text{trial}-\exsp|_\text{current})/k_{\rm B}T_\text{sa}]$. In the present study, approximately $2\times 10^{5}$ trial moves consisting of steps (i) and (ii) are carried out per cycle.   

After completion of a cycle, a lower $k_{\rm B}T_\text{sa}$ is chosen that  is 85\% the value of the previous simulated annealing temperature. A cycle at this lower temperature is then initiated unless  the run has converged according to the following metric: the difference in the minimum value of $-\exsp/k_{\rm B}$ obtained in four successive
temperature reduction cycles is  less than 0.1. 

\subsection{Dynamics simulations}
\label{sec:mol_dn_sim}

 Molecular dynamics simulations were carried out in the
microcanonical ensemble using the velocity-Verlet integration method with a time step of $0.001 \sigma \sqrt{m/\epsilon}$. The average temperature of all runs was $\epsilon/k_{\rm B}$, which was set by periodic velocity rescaling during pre-equilibration simulations.\cite{PhysRevE.51.4626} For comparison, a small number of Brownian (overdamped Langevin) simulations were also carried out using the Ermak algorithm using timestep $\Delta \text{t} = 0.002 \tau_\text{B}$ where $\tau_\text{B} = m D_0/\epsilon$ and $D_0 = 0.001 \sigma \sqrt{m/\epsilon}$.\cite{ErmakBrownian1975} A single tracer particle and $4000<N<8000$ hard-sphere-like fluid particles per (periodically-replicated) cubic simulation cell  of volume $V$ were used; the values of these parameters in a given simulation were chosen to realize a specified reduced density $\rho \sigma^3$ of the fluid. The tracer
diffusivity was computed by fitting the long time behavior of the mean
squared displacement to the Einstein relation for diffusion $\langle
\Delta \mathbf{r}^2 \rangle = 6 D t$.  Error bars represent 95\% confidence intervals determined by analyzing the results of between 30-192 independent trajectories.

\section{Results and discussion}
\label{sec:RandD}

\begin{figure}[h!]
        \centering
  \includegraphics{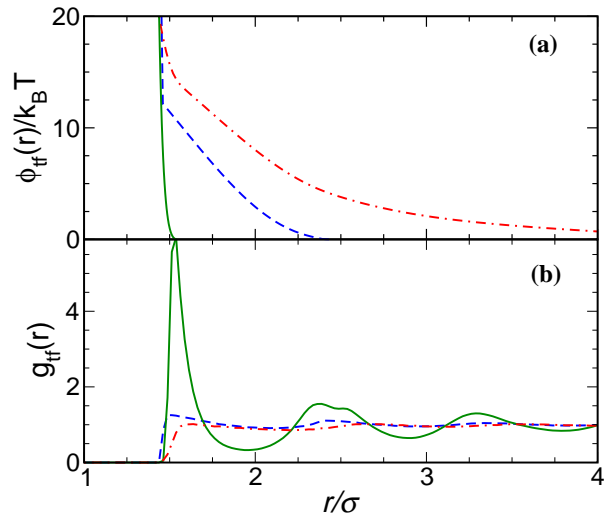}
  \caption{($a$) Potential [$\Vtp$] and ($b$) PRDF [$g_\text{tf}(r)$] between tracer particle of diameter $\sigma_\text{t}=2\sigma$
                 and fluid particles of diameter $\sigma$ and density $\rho \sigma^3=1$. Hard-sphere-like (green solid), 
                 $\stwo$-optimized (blue dashed), and $\exsp$-optimized (red dot-dashed) interactions are shown.}
  \label{fig:dens_wca_s2_sx}
\end{figure}

\begin{figure}[h!]
        \centering
  \includegraphics{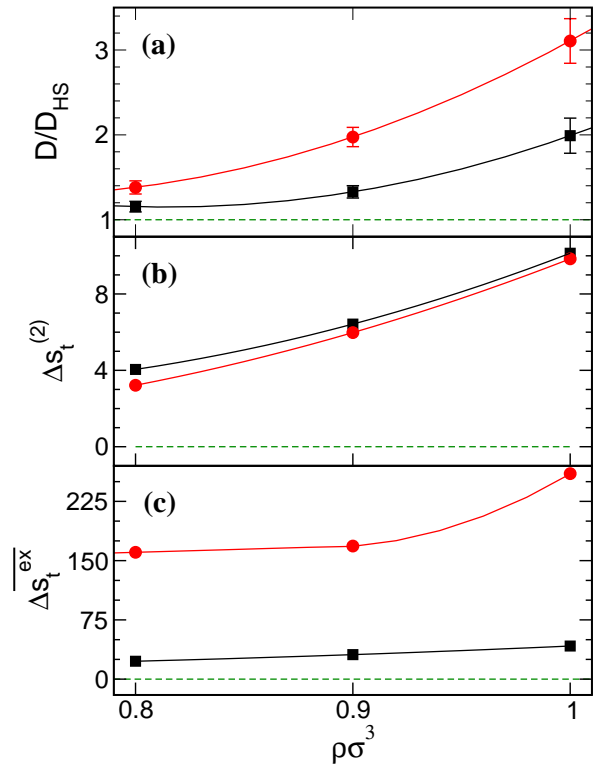}
  \caption{($a$) Ratio of diffusivity ($D$) for $\stwot$ - optimized (black square) and $\exsp$ - optimized (red circle) tracer to that 
                                        ($D_{HS}$) of a hard-sphere-like tracer with the same exclusion diameter ($\sigma_\text{t} = 2\sigma$) in a fluid of hard-sphere-like 
                                        particles of diameter $\sigma$ and density $\rho \sigma^3$. Enhancement of (b) $\stwot$ and (c) $\exsp$ (over 
                                        hard-sphere-like tracer values) due to optimization.}
  \label{fig:D_ratio_s2_sx_rho}
\end{figure}

In this section, we use methods described above to compare how the static structure and single-particle dynamics of an infinitely dilute hard-sphere-like tracer particle in a dense hard-sphere fluid compare to the same quantities for tracers whose hard-sphere interactions are augmented by additional pair interactions to optimize either $\stwot$ or $\exsp$. Except where indicated, all results for PRDFs and long-time diffusivities are obtained from molecular dynamics simulations.

As a starting point, Fig.~\ref{fig:dens_wca_s2_sx} shows pair potentials and PRDFs for the case of $\sigma_\text{t} = 2\sigma$ and $\rho\sigma^3 =1$. This is a moderately supercooled fluid; the thermodynamic freezing transition for hard spheres occurs at $\rho \sigma^3 \approx 0.94.$ The main feature to note in Fig.~\ref{fig:dens_wca_s2_sx}a is that the optimized tracer-fluid potentials differ considerably from the hard-sphere-like form due to the presence of longer-range soft repulsions. What effects should these repulsions have on tracer-particle diffusivity? On one hand, one might expect--based on arguments from hydrodynamics\cite{Einstein1905}--that the optimized tracers will show slower dynamics due to their larger effective size. On the other hand, Fig. 1b clearly illustrates that the main effect of the optimized potentials is to disrupt the coordination-shell ``cage'' structure surrounding the optimized tracers. Based on this disruption, we
expect the diffusivity to increase.

Fig. \ref{fig:D_ratio_s2_sx_rho}a shows that the diffusivity of tracers with optimized potentials has indeed increased by up to a factor of three (two) for the $\exsp$- $(\stwot$-) optimized potentials, respectively. For a hard-sphere tracer, increasing the fluid density results in more pronounced coordination-shell ``cages'' of surrounding fluid particles. The $\stwot$ and $\exsp$ optimization procedures work to destroy this caging structure at any density, which is why the structural and dynamic consequences of optimization are more pronounced in denser fluids. Fig.\ref{fig:D_ratio_s2_sx_rho}b illustrates that the difference between the two types of optimized potentials, clearly reflected in the tracer dynamics, does not manifest in $\stwot$. This is consistent with the observation that the shape of the pair correlation function alone cannot generally account for the dynamics of particles in dense fluids.\cite{BerthierTarjus2011} However, other aspects of static structure capture the difference. As is shown in Fig.\ref{fig:D_ratio_s2_sx_rho}c, when comparing tracers with the same hard-core diameter, those with higher diffusivities have considerably higher values of $\exsp$.

\begin{figure}[h]
        \centering
  \includegraphics{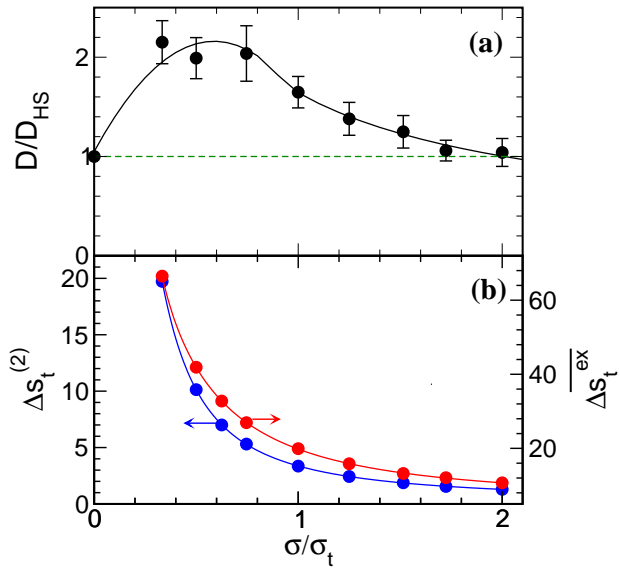}
  \caption{($a$) Ratio of diffusivity ($D$) for $\stwot$ - optimized tracer to that ($D_{HS}$) of a hard-sphere-like tracer 
                        with the same exclusion diameter, $\sigma_\text{t}$. Both tracers are in a fluid of hard-sphere-like particles 
                        of diameter $\sigma$ and density $\rho \sigma^3=1$.
  ($b$) The corresponding differences in $\stwot$ and $\exsp$ between the $\stwot$ - optimized tracer and hard-sphere-like tracer.}
  \label{fig:D_ratio_s2_sig}
\end{figure}

\begin{figure}[h]
        \centering
  \includegraphics{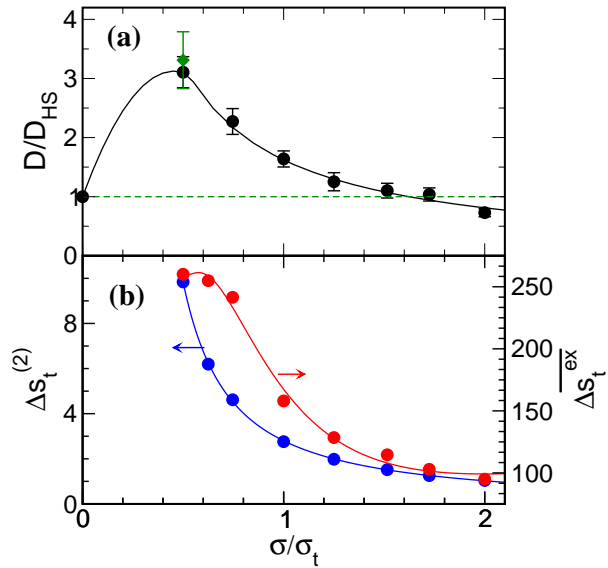}
  \caption{($a$) Ratio of diffusivity ($D$) for $\exsp$ - optimized tracer to that ($D_{HS}$) of a hard-sphere-like tracer 
                        with the same exclusion diameter, $\sigma_\text{t}$. Both tracers are in a fluid of hard-sphere-like particles 
                        of diameter $\sigma$ and density $\rho \sigma^3=1$. One data point (green diamond) is from Brownian dynamics simulations described
                        in the text.
                         ($b$) The corresponding differences in $\stwot$ and $\exsp$ between the $\exsp$ - optimized tracer and hard-sphere-like tracer. }
  \label{fig:D_ratio_sx_sig}
\end{figure}

The effect of $\stwot$- and $\exsp$-based optimizations for tracer particles with different bare hard-core diameters (relative to those of surrounding fluid particles) is explored in Figures \ref{fig:D_ratio_s2_sig} and \ref{fig:D_ratio_sx_sig}, respectively. For an infinitely large tracer, the first coordination shell of fluid particles constitutes an insignificant fraction of the tracer particle size. Any modification to this structure is expected to have a negligible effect on tracer diffusivity.\cite{Skinner2003} At the other extreme ($\sigma_\text{t}/\sigma = 0$), there is less structuring surrounding the point-like tracer to begin with, and thus the benefits of eliminating this structure through optimization are outweighed by the associated increase in effective particle size. Between these two limits, there is a maximum in the diffusivity ratio (as a function of tracer particle size) for $\stwot$ and $\exsp$ optimized potentials. In the next section, we introduce a model ``hard-sphere + Yukawa'' tracer to investigate the dynamic consequences of the trade-offs between effective particle size, softness of the interparticle potential, and fluid structure.


\subsection{Tracer-fluid interactions with increased $\stwot$ and $\exsp$:
  Mapping onto effective potentials}
\label{sec:interactions_faster_diffusion_experimental}
The form of the soft, repulsive potential resulting from optimizing $\exsp$ looks similar to a HCY \textbraceleft i.e., hard-core $\left[\phi^{WCA}(r)\right]$ + Yukawa $\left[\phi^Y(r)\right]$\textbraceright\ interaction, which is often used to approximately model particles with repulsive, screened-electrostatic forces. Separating the Yukawa contribution from the hard-sphere contribution allows us to tune in the soft repulsion  in a controlled way via the parameter $\alpha$, where $\alpha$ = 1 corresponds to a best-fit of the model to the $\exsp$-optimized potential for $\sigma_\text{t} = 2\sigma$ and  density $\rho \sigma^3=1$.
\begin{equation}
        \phi_{\rm tf }^{\rm{HCY}}(r) = \phi ^{WCA}(r) +\frac{\alpha\epsilon _Y\:\sigma}{r}\:e^{-\kappa (r-\sigma)}
\label{eq:hs_yuk}
\end{equation}
The parameters for the fit are $\epsilon_Y = 13.41 \epsilon$ and $\kappa \sigma = 0.82$.

\begin{figure}[h]
        \centering
  \includegraphics{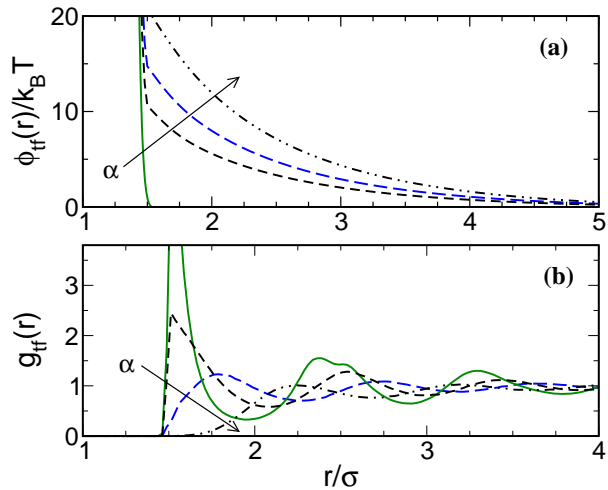}
  \caption{(($a$) Potential [$\Vtp$] and ($b$) PRDF [$g_\text{tf}(r)$] for  the HCY tracer particle with $\sigma_\text{t} = 2\sigma$ in a fluid of hard-sphere-like particles of diameter $\sigma$ and 
                                                        density $\rho \sigma^3=1$. Cases $\alpha=0$ (green solid),$\alpha=0.5$ (black dashed),  $\alpha=1$ (blue long-dash), and $\alpha=2$ (black dot-dashed) are shown.}
  \label{fig:yukawa_alpha_rdf}
\end{figure}

\begin{figure}[h]
        \centering
  \includegraphics{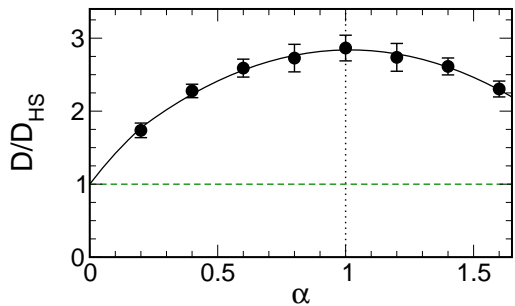}
  \caption{Ratio of diffusivity ($D$) for the HCY tracer to that ($D_{HS}$) of a hard-sphere-like tracer with the same 
                                                        exclusion diameter ($\sigma_\text{t} = 2\sigma$) in a fluid of hard-sphere-like particles of diameter $\sigma$ and 
                                                        density $\rho \sigma^3=1$.
}
  \label{fig:D_ratio_yukawa_alpha}
\end{figure}

Fig. \ref{fig:yukawa_alpha_rdf} shows HCY tracer pair potentials and PRDFs for various values of $\alpha$. Moving from $\alpha = 0$ to $\alpha = 1$, the main effect on the PRDF is to decrease the height of the first peak and reduce the structuring in more distant coordination shells. Fig. \ref{fig:D_ratio_yukawa_alpha}a shows a corresponding enhancement of the tracer diffusivity, which might be expected given that the potential and structural changes
resemble those of the $\exsp$-based optimization process.  

However, for $\alpha > 1$, the model tracer-fluid potential exhibits a Yukawa repulsion that is stronger than optimal. In fact, as can be seen from the PRDF in Fig. \ref{fig:yukawa_alpha_rdf}b, the resulting tracer exclusion core [separations for which $g_{\rm tf} (r) \approx 0$] becomes noticeably larger than that of the underlying WCA potential. Accordingly, Fig. \ref{fig:D_ratio_yukawa_alpha} shows that the tracer diffusivity decreases with increasing $\alpha$ for $\alpha > 1$, as should be expected since the main effect in this range is to increase the effective hard-core diameter. Note that such potentials with different hard-core diameters than the underlying WCA potential are avoided by construction in the $\exsp$-based optimization.

\section{Conclusions}
\label{sec:conclusion}

Using a combination of simulations and liquid-state theory, we have shown that the long-time diffusivity of a hard-sphere-like tracer particle with a given exclusion diameter in a dense hard-sphere fluid can be significantly enhanced by adding a soft repulsion to its interactions with neighboring particles. An important effect of this repulsion is to disrupt the coordination shells that otherwise surround the bare hard-sphere tracer. The form of the required repulsions can be determined theoretically by maximizing either the tracer's partial molar excess entropy or its partial radial-distribution function contribution to the fluid's excess entropy. We show that the long-time diffusivities of tracers with $\stwot$- and $\exsp$-optimized interactions can be higher than those of the corresponding hard-sphere tracers by more than factor of two and three, respectively, for moderatately supercooled liquid-state points.

In future studies, we plan to investigate the effect of the aforementioned soft repulsions on the position-dependent diffusivity of neighboring particles. Such studies might help to understand whether the effects shown here are consistent with an enhanced mobility (or reduced viscosity) of the surrounding fluid in the neighborhood of the tracer. We also plan to explore the possibility of optimizing tracer dynamics in deeply supercooled liquids, where the caging structure (and its dynamical consequences) are significantly more pronounced. 


T.M.T. acknowledges support of the
Welch Foundation (F-1696) and the National Science Foundation (CBET-1065357).
J. R. E. acknowledges financial support of the National Science Foundation (CBET-0828979). The Texas
Advanced Computing Center (TACC) provided computational resources for
this study.


\newpage 
\footnotesize{
\providecommand*{\mcitethebibliography}{\thebibliography}
\csname @ifundefined\endcsname{endmcitethebibliography}
{\let\endmcitethebibliography\endthebibliography}{}

}
\end{document}